\documentclass[reprint,amsmath,amssymb, aps]{revtex4-1}
\usepackage{graphicx}
\usepackage{color}
\usepackage{amsmath}
\usepackage{amssymb}
\usepackage{amsthm}
\usepackage{booktabs}
\usepackage{float}
\usepackage[dvipdfm,colorlinks,urlcolor=blue,linkcolor=blue,anchorcolor=blue,citecolor=blue]{hyperref}
\usepackage{lineno}

\begin{document}

\title{Thermalization in asymmetric harmonic chains}

\author{Weicheng Fu$^{1,2,3}$}
\email{fuweicheng@tsnu.edu.cn}
\author{Sihan Feng$^4$}
\author{Yong Zhang$^{4,3}$}
\email{yzhang75@xmu.edu.cn}
\author{Hong Zhao$^{4,3}$}
\email{zhaoh@xmu.edu.cn}
\affiliation{
$^1$ Department of Physics, Tianshui Normal University, Tianshui 741001, Gansu, China\\
$^2$ Key Laboratory of Atomic and Molecular Physics $\&$ Functional Material of Gansu Province, College of Physics and Electronic Engineering, Northwest Normal University, Lanzhou 730070, China\\
$^3$ Lanzhou Center for Theoretical Physics, Lanzhou University, Lanzhou, Gansu 730000, China\\
$^4$ Department of Physics, Xiamen University, Xiamen 361005, Fujian, China
}

\date{\today }

\begin{abstract}
The symmetry of the interparticle interaction potential (IIP) plays a critical role in determining the thermodynamic and transport properties of solids. This study investigates the isolated effect of IIP asymmetry on thermalization. Asymmetry and nonlinearity are typically intertwined. To isolate the effect of asymmetry, we introduce a one-dimensional asymmetric harmonic (AH) model whose IIP possesses asymmetry but no nonlinearity, evidenced by energy-independent vibrational frequencies. Extensive numerical simulations confirm a power-law relationship between thermalization time ($T_{\rm eq}$) and perturbation strength for the AH chain, revealing an exponent larger than the previously observed inverse-square law in the thermodynamic limit. Upon adding symmetric quartic nonlinearity into the AH model, we systematically study thermalization under combined asymmetry and nonlinearity. Matthiessen's rule provides a good estimate of $T_{\rm eq}$ in this case. Our results demonstrate that asymmetry plays a distinct role in enhancing higher-order effects and governing relaxation dynamics.
\end{abstract}

\maketitle

\section{Introduction}
In the 1950s, Fermi, in collaboration with Pasta, Ulam, and Tsingou (FPUT), conducted the first numerical experiments to test the ergodic hypothesis using a simple mechanical system of springs and masses \citep{Fermi1955}. Their pioneering work unexpectedly revealed that the system, far from equilibrium, did not evolve into the anticipated thermalized state but instead returned to a state of near non-equilibrium, a phenomenon now known as the FPUT recurrences. This finding sparked extensive research aimed at explaining and understanding this phenomenon through longer simulations and larger system sizes in various nonlinear chains \citep{Chaos2005, 2008LNP728G, ferguson1982nonlinear, casetti1997fermi, Ponno2011Chaos, Benettin2013JSP, benettin2018fermi, goldfriend2019equilibration, grava2020adiabatic, benettin2020understanding}.

In recent years, the application of wave turbulence theory has significantly advanced understanding of this problem \citep{Onorato2015PNAS, PhysRevLett.120.144301, Pistone2018EPL, PhysRevLett.129.063901, ONORATO20231, ferraro2025}. Extensive numerical simulations of various models have revealed that, in the thermodynamic limit, there exists a universal scaling law for the thermalization behavior of near-integrable systems: thermalization time ($T_{\rm eq}$) is inversely proportional to the square of perturbation strength \citep{Fu_2019njp, Fu_2019PRER, Pistone2019ME, Fu2019diatomic,Feng_2022, wang2020wave, fu2021effect,Wang_2024}. Recent studies indicate that this inverse-square law holds true even in high-dimensional lattice systems, unaffected by lattice structures, interaction potentials, or whether the lattice being ordered or not \citep{PhysRevLett.132.217102}. To observe this universal law, a suitable reference integrable system must be chosen to define the perturbation strength, ensuring an accurate description of the system's ability to thermalize. For example, a general nonlinear monatomic chain can be considered as a perturbation of the Toda model, with the perturbation strength defined as the distance from the Toda system \citep{Fu_2019njp}. When the Toda integrability is broken, such as in diatomic chains \citep{Fu2019diatomic,Feng_2022}, mass-disordered chains \citep{wang2020wave, fu2021effect}, chains with on-site potentials \citep{Pistone2018EPL}, or high-dimensional systems \citep{PhysRevLett.132.217102}, the system is considered a perturbation of the harmonic one, with perturbation strength defined relative to the harmonic reference point.

However, deviations from this universal law, such as steeper slopes, have been observed in chains with cubic or quintic nonlinearity \citep{Fu_2019PRER, Pistone2019ME}. Two primary explanations for this deviation exist: one suggests a finite-size effect (discreteness) \citep{Pistone2019ME}, while the other points to higher-order effects \citep{Fu_2019PRER}. Notably, both third- and fifth-order nonlinearities involve asymmetric interaction potentials, meaning that the amplitude of forces corresponding to the same displacement is not equal in the tension and compression states.

Asymmetric interparticle interaction potentials (IIPs) play a critical role in lattice models, influencing thermal expansion effects, which symmetric potentials cannot produce \citep{kittel1996introduction}. Furthermore, IIP asymmetry significantly affects transport properties \citep{lepri2016thermal}. For instance, in a 1D momentum-conserving system, the bulk viscosity of a system with symmetric IIP remains finite in the thermodynamic limit, while it diverges for an asymmetric IIP \citep{PhysRevE.72.031202, lee2008detailed, PhysRevE.73.060201, Delfini_2007}. When the IIP is symmetric, the heat conductivity ($\kappa$) scales with system size as $\kappa \sim N^{1/2}$ \cite{Delfini_2007,PhysRevLett.108.180601}, although mode coupling theory \cite{Lepri1998, PhysRevLett.92.074302}, Peierls-Boltzmann kinetic theory \citep{PhysRevE.68.056124}, and wave turbulence theory \citep{PhysRevLett.125.024101} predict $\kappa \sim N^{2/5}$, which is supported by recent ultra-large-scale nonequilibrium simulations \citep{2024JPSJ.93.053001}. In contrast, for asymmetric IIP, $\kappa \sim N^{1/3}$ \citep{PhysRevLett.108.180601, spohn2014nonlinear, PhysRevE.72.031202, Delfini_2007,SSPMA_zhao,Luo2025}. More surprisingly, normal heat conduction (i.e., $\kappa$ independent of system size in the thermodynamic limit) has been observed in chains with asymmetric IIP in the near-integrable regime \citep{PhysRevE.85.060102, Zhong_2013, chen2016key, Jiang_2016}, challenging conventional views. While some researchers attribute this to finite-size effects \citep{PhysRevE.88.052112, Dhar2014JSP}, it has been shown that systems with asymmetric IIP exhibit a larger kinetic region \citep{PhysRevE.90.032134, chen2016key, Jiang_2016, PhysRevE.97.010103}, suggesting that asymmetric IIP may lead to diffusive kinetic behaviors, whereas symmetric IIP does not \citep{lepri2020too}.

This raises the question of whether asymmetric IIP plays a distinct role in thermalization. In the models discussed above, such as those with odd-order nonlinearity, the effects of asymmetry and nonlinearity are intertwined, making it challenging to separate their contributions. To address this, we introduce an asymmetric harmonic (AH) model, which is purely asymmetric and not nonlinear (since nonlinearity typically involves the frequency of the motion of particles depends on the amplitude or, equivalently, the input energy \citep{Campbell2004}) to investigate the influence of pure asymmetry on thermalization. Subsequently, we introduce quartic nonlinearity into the AH model to explore the thermalization behavior when asymmetry and nonlinearity are interwoven. In the following sections, we first present the models and methods in Sec.~\ref{sec2}, followed by the numerical results in Sec.~\ref{sec3}, and conclude with a summary and discussion in Sec.~\ref{sec4}.

\section{Models and method}\label{sec2}

We consider a homogeneous lattice consisting of $N$ particles of unit mass, with the Hamiltonian given by
 \begin{equation}\label{eq:1}
  H = \sum_{j=1}^N\frac{p_j^2}{2}+\sum_{j=0}^NV(q_{j+1}-q_j),
 \end{equation}
where $p_j$ and $q_j$ represent the momentum and displacement from the equilibrium position of the $j$-th particle, respectively, and $V$ is the nearest-neighbor interaction potential. To investigate the effects of asymmetry and nonlinearity on thermalization, we consider two types of interaction potentials. The first is the AH potential \citep{Zhong_2013}, defined as
\begin{equation}\label{eq:2}
  V_{\text{AH}}(x) = \frac{1}{2}
  \begin{cases}
   (1-r)x^2,&x<0;\\
   (1+r)x^2,& \text{otherwise},
  \end{cases}~r\in[0,1),
\end{equation}
where $r$ is a free parameter that controls the degree of asymmetry. Actually, Eq. (\ref{eq:2}) can be rewritten as
\begin{equation}\label{eq:2b}
  V_{\text{AH}}(x) = \frac{1}{2}x^2+\text{sgn}(x)\frac{r}{2}x^2,
\end{equation}
where $\text{sgn}(x)$ is the sign function. It is clearly shown that $r$ also is \emph{the perturbation strength} relative to the harmonic (reference integrable) system. The AH potential is similar to the broken linear potential studied in the original FPUT work \citep{Fermi1955}, which corresponds to a piecewise-smooth dynamical system \cite{Bernardo2008}. In practice, non-smooth dynamics arising from switches, impacts, sliding, and abrupt changes are common in various fields of physics, biology, and engineering \cite{10.1063/5.0138169}.

The AH model has the advantage that its dynamics are independent of the system's energy (temperature) and depend solely on the parameters $r$ and $N$ \citep{Zhong_2013}. We next introduce a symmetric fourth-order nonlinearity into the AH potential, yielding the AH-$\beta$ potential
 \begin{equation}\label{eq:3}
  V_{\text{AH-}\beta}(x) = V_{\text{AH}}(x)+\frac{\beta}{4}x^4,
 \end{equation}
where $\beta$ is a positive parameter controlling the strength of nonlinearity. Note that when $r = 0$, the AH-$\beta$ potential reduces to the FPUT-$\beta$ potential. The dimensionless parameter $\tilde{\beta} = \beta \varepsilon$ governs the strength of nonlinearity for the AH-$\beta$ model, where $\varepsilon = E/N$ is the energy density, with $E$ being the total energy of the system. For simplicity, we omit the tilde in subsequent expressions.

In this work, we consider fixed boundary conditions, i.e., $q_0 = q_{N+1} = 0$. The normal modes of the chain are defined by
\begin{align}\label{eq:4}
  \begin{cases}
   Q_k&=\sqrt{\frac{2}{N+1}}\sum_{j=1}^Nq_j\sin\left(\frac{jk\pi}{N+1}\right),\\
   P_k&=\sqrt{\frac{2}{N+1}}\sum_{j=1}^Np_j\sin\left(\frac{jk\pi}{N+1}\right).
  \end{cases}
\end{align}
where $k = 1, 2, \dots, N$ are the mode indices. The frequency $\omega_k$ and energy $E_k$ of the $k$-th mode are given by
\begin{equation}\label{eq:5}
  \omega_k=2\sin\left(\frac{k\pi}{2N+2}\right),\quad E_k=\frac{1}{2}\left(P_k^2+\omega_k^2Q_k^2\right).
\end{equation}
To each mode $k$, we associate a phase $\varphi_k$ defined by
\begin{align}\label{eq:7}
  \begin{cases}
   Q_k&=\sqrt{\frac{2E_k}{\omega_k^2}}\sin{\left(\varphi_k\right)},\\
   P_k&=\sqrt{2E_k}\cos{\left(\varphi_k\right)}.
  \end{cases}
\end{align}
In the thermalized state, energy equipartition is achieved, meaning
\begin{equation}\label{eq:8}
  \lim_{T\rightarrow\infty}\bar{E}_k(T)\simeq\varepsilon, \quad k=1,~\dots,~N,
\end{equation}
where $\bar{E}_k(T)$ represents the time average of $E_k$ over the time interval $[\theta T, T]$, with $\theta \in [0, 1)$ controlling the time average window, i.e.,
\begin{equation}\label{eq:10}
  \bar{E}_k(T)=\frac{1}{(1-\theta)T}\int_{\theta T}^TE_k(P(t),Q(t))dt.
\end{equation}
For numerical simulations, we use $\theta = 2/3$, which accelerates the calculations and reduces memory effects from the initial state, as noted in Ref.~\citep{JStatPhys2011}.

To measure how close the system is to thermal equilibrium, we introduce a modified version of the normalized effective relative number of degrees of freedom \citep{PhysRevA.31.1039}, denoted $\xi(t)$, which is sensitive to the early energy growth of high-frequency modes ($k \geq N/2$), as
\begin{equation}\label{eqXi}
  \xi(t)=\tilde{\xi}(t)\frac{e^{\eta(t)}}{N/2},
\end{equation}
where $\eta(t)$ is the spectral entropy, defined by
 \begin{equation}\label{eq:12}
  \eta(t)=-\sum_{k=N/2}^{N}w_k(t)\log[w_k(t)],
 \end{equation}
with
\begin{equation}\label{eq:13}
  w_k(t)=\frac{\bar{E}_k(t)}{\sum_{l=N/2}^{N}\bar{E}_l(t)},\quad \tilde{\xi}(t)=\frac{\sum_{k=N/2}^N\bar{E}_k(t)}{\frac{1}{2}\sum_{l=1}^N\bar{E}_l(t)}.
\end{equation}
As the system approaches thermal equilibrium, $\xi$ saturates at the value 1.

In our numerical simulations, the equations of motion are integrated using the eighth-order Yoshida algorithm \citep{YOSHIDA1990262} with a typical time step of $\Delta t = 0.05$. The relative error in energy conservation is less than $10^{-5}$, and further reducing the time step by an order of magnitude does not yield significant differences. To suppress fluctuations, we average over 120 random choices of initial phases uniformly distributed in $[0, 2\pi]$. Energy is initially distributed among $10\%$ of the lowest-frequency modes ($0<k/N\leq0.1$) in all simulations, and we have verified that varying the percentage of excited modes does not produce qualitative differences. In this study, the energy density is kept constant at $10^{-3}$, i.e., $\varepsilon=10^{-3}$.

\section{Numerical Results}\label{sec3}

\begin{figure*}[t]
 \centering
 \includegraphics[width=1.8\columnwidth]{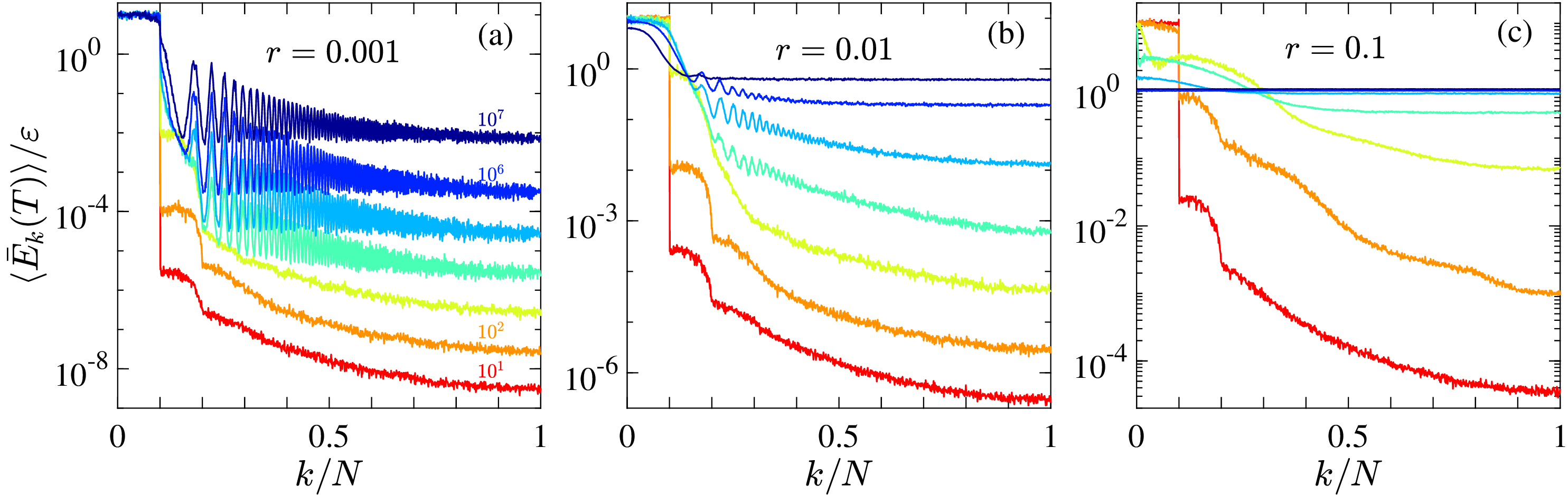}\\
 \caption{(a)-(c) Time evolution of the normalized modal energy spectrum $\langle \bar{E}_k(t)/\varepsilon \rangle$ as a function of normalized wave number $k/N$ for the AH chain under different perturbation strengths $r$, plotted on a semi-logarithmic scale. In each panel, curves from bottom to top  correspond to times $T = 10^1, 10^2, 10^3, 10^4, 10^5, 10^6, 10^7$, respectively. All simulations are performed for a chain of size $N = 1023$, with initial excitations restricted to modes in the range $0 < k/N \leq 0.1$.
 }\label{fig1}
\end{figure*}

Figure~\ref{fig1} presents the numerical results of $\langle \bar{E}_k(t)/\varepsilon \rangle$ as a function of $k/N$ at various selected times for the AH chain with different perturbation strengths $r$, and with fixed system size $N = 1023$. The results show that the energy initially concentrated in the excited modes gradually redistributes to the other modes over time, with the energy of the remaining modes increasing continuously. This behavior contrasts with that observed in the FPUT model~\citep{Benettin2009JSP,JStatPhys2011,Benettin2013JSP} and the perturbed Toda model~\citep{Fu_2019njp}, where $\langle \bar{E}_k(t)/\varepsilon \rangle$ maintains an exponentially distributed profile over a large initial time scale, corresponding to the so-called metastable state. Due to the non-smoothness of the AH model at $x = 0$, it lacks Toda integrability, and thus no metastable state is observed. As shown in Figs.~\ref{fig1}(a) to \ref{fig1}(c), the system reaches equipartition more rapidly as $r$ increases.

\begin{figure}[t]
 \centering
 \includegraphics[width=1\columnwidth]{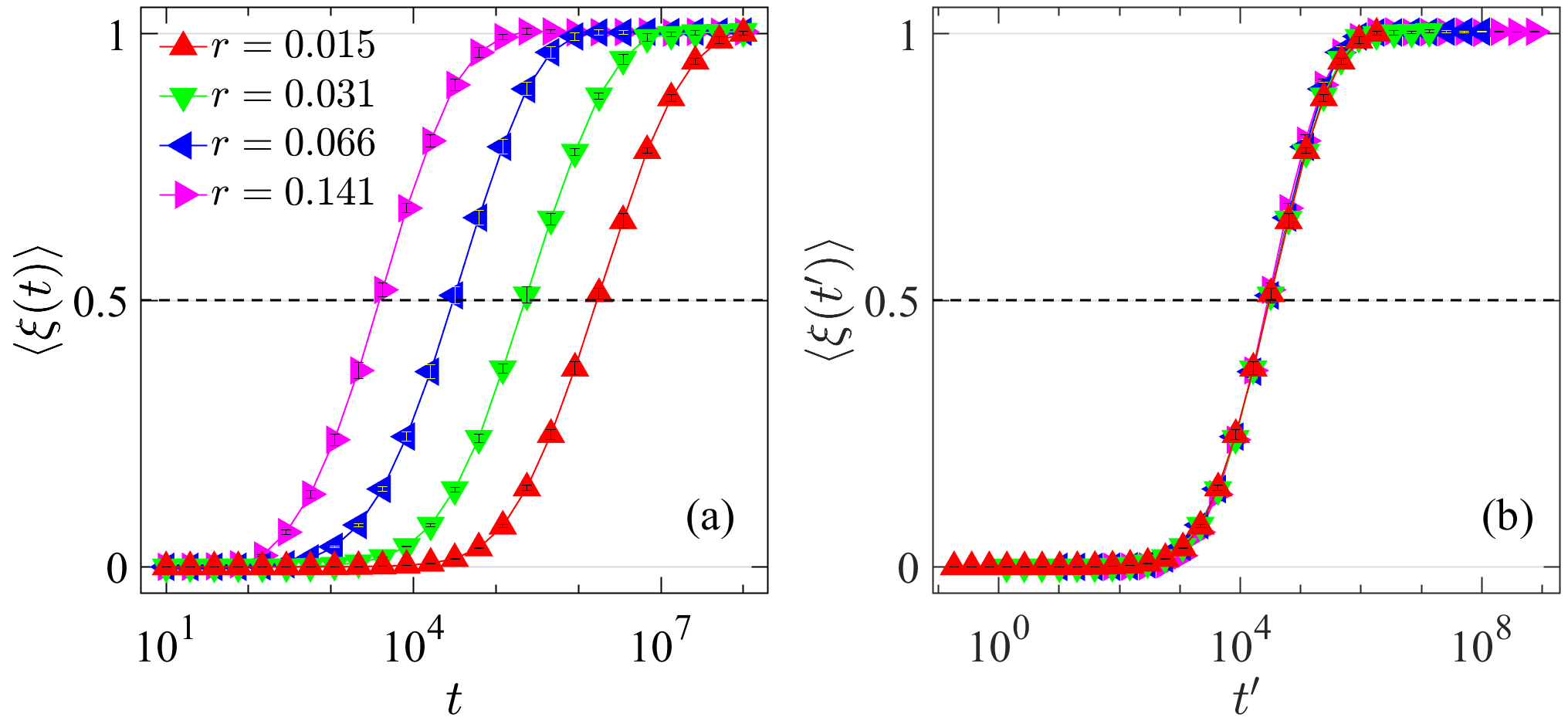}
 \caption{(a) The function of $\langle \xi(t) \rangle$ versus $t$ for the AH chain with different perturbation strength $r$. (b) Same data as in (a), but with the curves horizontally shifted ($r=0.066$ unshifted) to achieve complete overlap, highlighting scaling behavior. In all cases, $N=2047$ is kept fixed.}\label{fig2}
\end{figure}

To observe the overall thermalization dynamics and determine the thermalization time, we examine the evolution of $\langle \xi(t) \rangle$, as defined in Eq.~(\ref{eqXi}). Figure~\ref{fig2} presents the numerical results for the AH chain with different values of $r$. It is evident that, over a sufficiently large time scale, all $\langle \xi(t) \rangle$ values increase from 0 to 1, following very similar sigmoidal profiles, indicating that energy equipartition is eventually achieved. Additionally, the time required to reach the equipartition state increases as $r$ decreases.

Physically, the thermalization time $T_{\rm eq}$ is defined as the time at which $\langle \xi(t) \rangle$ reaches the threshold value of 1. However, for practical purposes, we are generally concerned with the scaling behavior of $T_{\rm eq}$, rather than its exact value. To reduce computational cost, we define $T_{\rm eq}$ as the time when $\langle \xi(t) \rangle$ reaches the threshold value of 0.5. While this is an arbitrary choice, it does not affect the scaling behavior of $T_{\rm eq}$ \citep{2008LNP728G}. As shown in Fig.~\ref{fig2}(b), the sigmoidal profiles in Fig.~\ref{fig2}(a) can be made to overlap completely upon suitable shifts, confirming that the specific threshold value does not influence the scaling law of $T_{\rm eq}$. In the following, we will explore how $T_{\rm eq}$ depends on $r$ and $N$ for the AH chain, and on $r$ and $\beta$ for the AH-$\beta$ chain.

\begin{figure*}[t]
 \centering
 \includegraphics[width=1\textwidth]{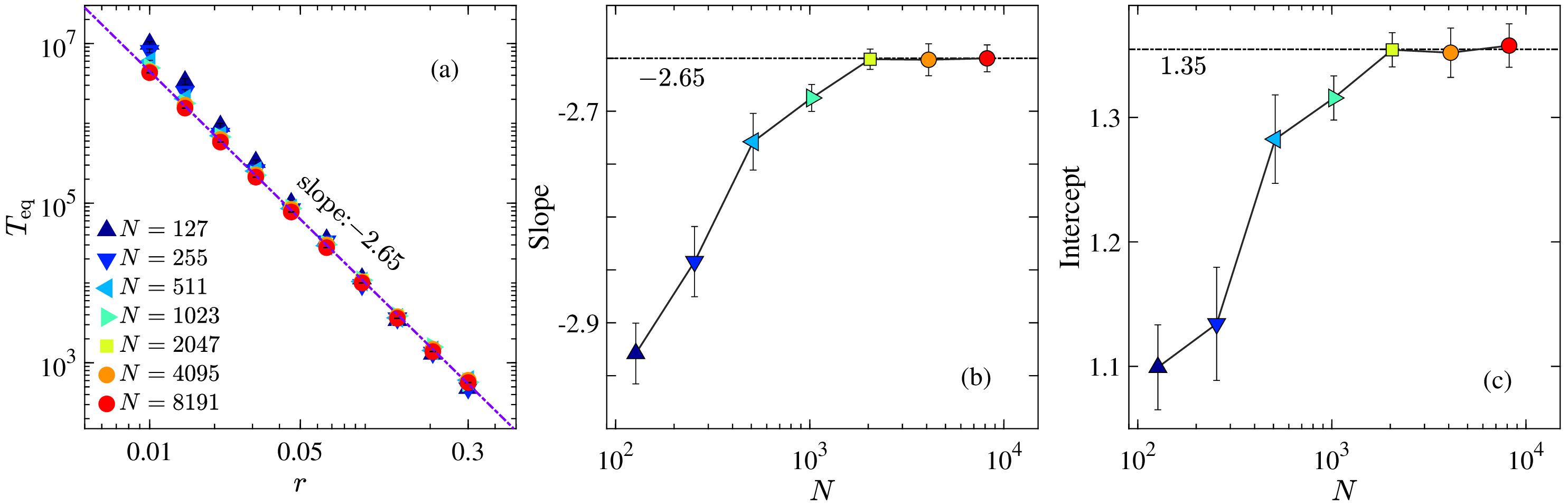}
 \caption{(a) Thermalization time $T_{\rm eq}$ as a function of perturbation strength $r$ for the AH chain with various system sizes, plotted on a log-log scale. The dashed reference line indicates a power-law slope of $-2.65$. (b) Slope and (c) intercept of the linear fits in panel (a) as functions of system size, shown on a semi-logarithmic scale.
}\label{fig3}
\end{figure*}

In Fig.~\ref{fig3}(a), we show the dependence of $T_{\rm eq}$ on $r$ for the AH chain with various values of $N$, on a log-log scale. For the range of $r$ explored, $T_{\rm eq}$ becomes nearly independent of $N$ as $N$ increases further. The numerical results suggest that $N = 2047$ is sufficiently large for the thermodynamic limit to be effectively reached. It is observed that $T_{\rm eq}$ versus $r$ follows a power-law behavior,
\begin{equation}
  T_{\rm eq}\propto r^{\lambda}.
\end{equation}
Figures~\ref{fig3}(b) and \ref{fig3}(c) show, respectively, the dependence of the slope $\lambda$ and the intercept of the linear fitting for the data in Fig. \ref{fig3}(a) on $N$. It can be seen that $\lambda$ quickly saturates at $-2.65$, while the intercept stabilizes at $1.35$. From this, we can estimate the thermalization time for the AH model in the thermodynamic limit as
\begin{equation}\label{eqTeqAH}
  T_{\rm eq}^{\text{AH}}\simeq 10^{1.35}r^{-2.65}\simeq22.39r^{-2.65},
\end{equation}
which deviates from the previously observed inverse-square law \citep{Fu_2019njp, Fu_2019PRER, Pistone2019ME, Fu2019diatomic, wang2020wave, fu2021effect,Feng_2022}. This phenomenon of a steeper slope is also observed in models with odd-order nonlinearity (e.g., the FPUT-$\alpha$ chain) \citep{Fu_2019PRER, Pistone2019ME}. The mechanism behind this deviation remains unclear. Ref. \citep{Pistone2019ME} suggests that the deviation is due to finite-size effects (discreteness), while Ref. \citep{Fu_2019PRER} shows that the results for different system sizes nearly coincide within the parameter range studied, yet the steeper slope persists. Additionally, it has been pointed out that asymmetric IIP leads to asymmetric spectral peaks of modes (i.e. a higher-order effect), which is considered the root cause of the deviation. Despite the AH model being purely asymmetric and devoid of nonlinearity, it still exhibits this deviation, further confirming that asymmetry results in a steeper slope. In other words, the asymmetry of the IIP enhances the contribution of higher-order effects. Next, we will investigate the role of nonlinearity, exemplified by the AH-$\beta$ model [see again Eq.~(\ref{eq:3})].

\begin{figure}[t]
 \centering
 \includegraphics[width=1\columnwidth]{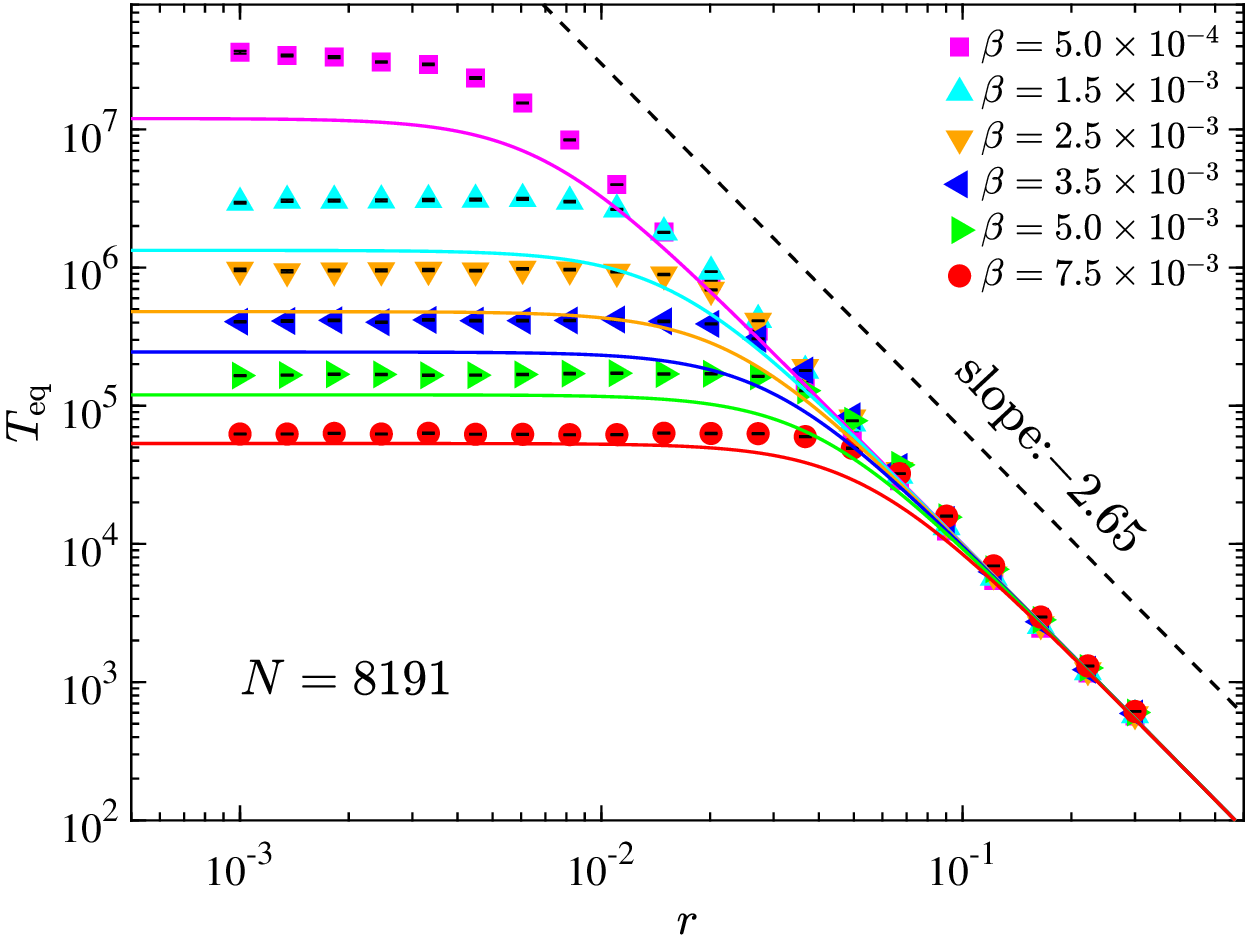}
 \caption{The thermalization time $T_{\rm eq}$ as a function of the perturbation strength $r$ for the AH-$\beta$ chain with different strength of nonlinearity $\beta$ at fixed system size. The solid curves are estimations given by Eq. (\ref{eqTeqAHB}). The dashed line serves as reference. Initial excitation of modes with $0<k/N\leq0.1$.}\label{fig4}
\end{figure}

Figure~\ref{fig4} presents the dependence of $T_{\rm eq}$ on $r$ in log-log scale for the AH-$\beta$ chain at fixed system size $N=8191$, under various values of $\beta$. As $r$ decreases, $T_{\rm eq}$ approaches a saturation value, and this value declines with increasing $\beta$. This behavior indicates that, in the small-$r$ regime, thermalization is predominantly governed by the quartic nonlinearity, whereas for large $r$, the asymmetry becomes the dominant factor. These observations suggest a competition between the asymmetric potential and the quartic nonlinearity in driving thermalization. Assuming their contributions are independent, as proposed in Ref.~\citep{Fu_2019PRER}, Matthiessen's rule (MR) \citep{Srivastava:215326} can be applied to estimate the total thermalization time, that is,
 \begin{equation}\label{eqTeqAHB0}
 \frac{1}{T_{\rm eq}^{\text{AH-}\beta}} = \frac{1}{T_{\rm eq}^{\text{AH}}} + \frac{1}{T_{\rm eq}^{\text{FPUT-}\beta}},
 \end{equation}
 where $T_{\rm eq}^{\text{AH}}$ is given by Eq.~(\ref{eqTeqAH}), and $T_{\rm eq}^{\text{FPUT-}\beta} \propto \beta^{-2}$ in the thermodynamic limit \citep{Fu_2019PRER, Pistone2019ME}. Based on Ref.~\citep{Fu_2019PRER}, we adopt the empirical estimate
 \begin{equation}\label{eqTeqFPUTB}
 T_{\rm eq}^{\text{FPUT-}\beta} \simeq 3\beta^{-2}.
 \end{equation}
 Combining Eqs.~(\ref{eqTeqAH}), (\ref{eqTeqAHB0}), and (\ref{eqTeqFPUTB}) yields an explicit expression for $T_{\rm eq}$ in the AH-$\beta$ chain
 \begin{equation}\label{eqTeqAHB}
 T_{\rm eq}^{\text{AH-}\beta} = \frac{67.17 r^{-2.65} \beta^{-2}}{22.39 r^{-2.65} + 3\beta^{-2}}.
 \end{equation}

The solid lines in Fig.~\ref{fig4} correspond to the prediction of Eq.~(\ref{eqTeqAHB}), showing good agreement with numerical data. However, the deviation increases as $\beta$ decreases, which is attributed to finite-size effects. Prior works~\citep{PhysRevLett.120.144301, Onorato2015PNAS} have shown that exact four-wave resonances, responsible for $T_{\rm eq} \propto \beta^{-2}$, require large system sizes ($N \geq 163264$). Nonetheless, due to nonlinearity-induced spectral broadening, quasi-resonant four-wave interactions can still occur in finite systems~\citep{PhysRevLett.95.264302,PhysRevE.111.024122}, especially at stronger nonlinearities. Thus, the $\beta^{-2}$ scaling emerges only for either large systems under weak nonlinearity or moderate systems under stronger nonlinearity~\citep{Fu_2019PRER}, explaining the discrepancy in the small-$\beta$ regime and its disappearance as $\beta$ increases.
Furthermore, Eq.~(\ref{eqTeqAHB}) implies that $T_{\rm eq}$ decreases monotonically with both $r$ and $\beta$ since $\partial T_{\rm eq} / \partial r < 0$ and $\partial T_{\rm eq} / \partial \beta < 0$ for fixed $\beta$ and $r$, respectively. However, the actual system behavior may exhibit deviations from this monotonic trend due to intertwined interactions.

\begin{figure}[t]
 \centering
 \includegraphics[width=1\columnwidth]{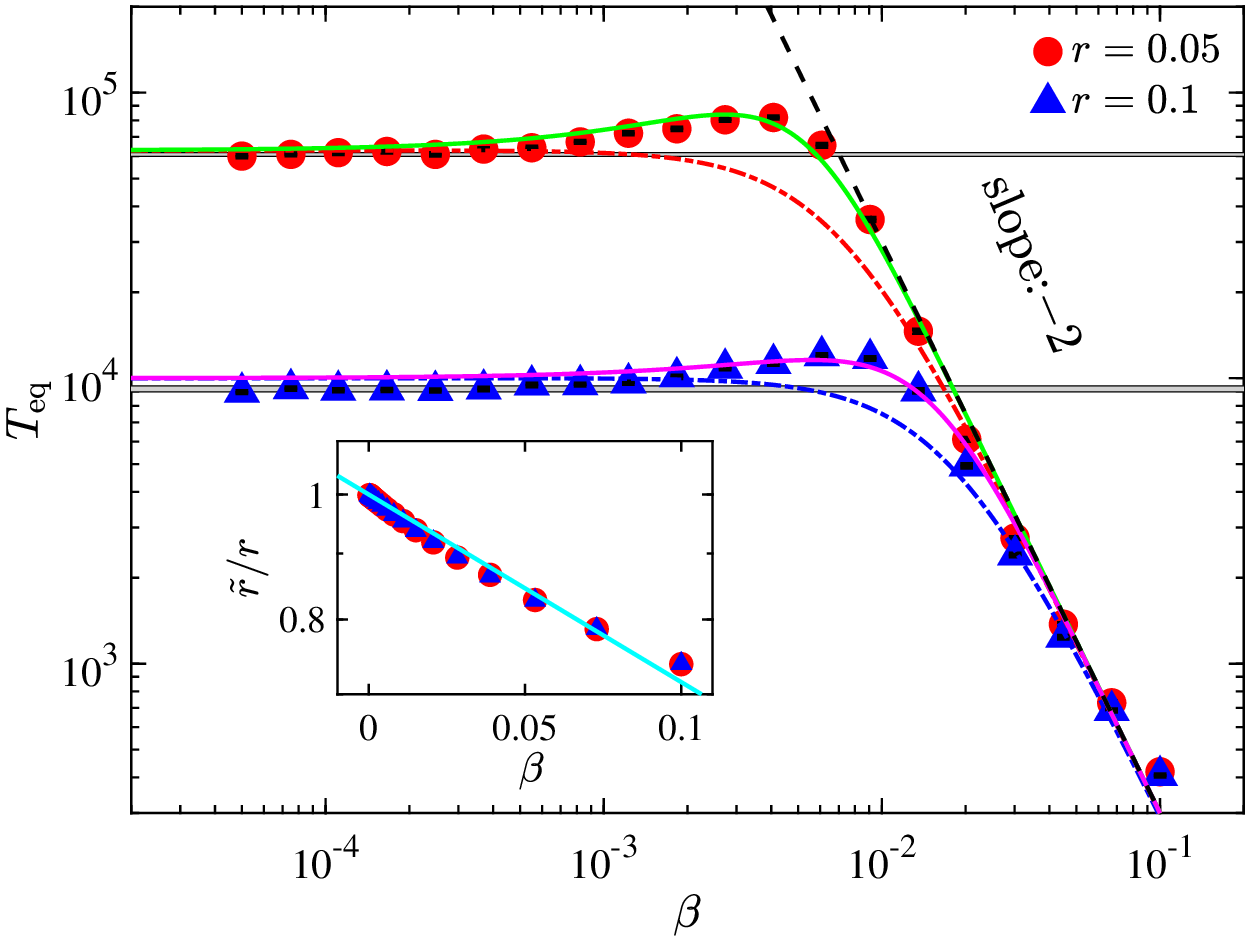}
 \caption{
 Thermalization time $T_{\rm eq}$ as a function of nonlinearity strength $\beta$ for the AH-$\beta$ chain, with fixed perturbation strength $r = 0.05$ (red circles) and $r = 0.1$ (blue up-triangles). The dashed-dotted curves represent theoretical predictions from Eq.~(\ref{eqTeqAHB}), while the solid lines are fitting curves based on the modified expression (\ref{eqTeqAHBM}). The dashed line corresponds to the estimate from Eq.~(\ref{eqTeqFPUTB}), shown for reference. Fitting parameters: $\gamma = 72.21$ for $r = 0.05$; $\gamma = 18.58$ for $r = 0.1$. The horizontal gray band denotes the $T_{\rm eq}$ of the AH model (i.e., $\beta = 0$), with its width indicating the numerical uncertainty. All simulations are performed with fixed system size $N = 2047$.
Inset: Normalized effective asymmetry $\tilde{r}/r$ as a function of $\beta$, plotted on a semi-logarithmic scale (log scale on the vertical axis). Data points correspond to the same parameters as in the main panel. The solid reference line shows an exponential decay $e^{-3.35\beta}$.
 }\label{fig5}
\end{figure}

Figure~\ref{fig5} shows the thermalization time $T_{\rm eq}$ as a function of $\beta$ on a log-log scale for the AH-$\beta$ chain at fixed system size $N = 2047$, with $r = 0.05$ (red circles) and $r = 0.1$ (blue triangles). The dashed-dotted lines represent theoretical predictions from Eq.~(\ref{eqTeqAHB}). While the general trend of the theoretical curves aligns with the simulation data, notable deviations emerge in the intermediate $\beta$ regime. Specifically, $T_{\rm eq}$ exhibits a non-monotonic dependence on $\beta$, i.e., it initially increases with increasing nonlinearity, then decreases, contrary to the common expectation that stronger nonlinearity facilitates thermalization.
This discrepancy raises questions about the applicability of MR in this context. In condensed matter theory, MR assumes that scattering mechanisms are independent. However, the fourth-order nonlinearity appears to suppress the effective asymmetry of the IIP, thereby invalidating this assumption. As discussed in Ref.~\citep{fu2015effective}, in the limit $\beta \to \infty$, the potential becomes effectively symmetric and the effective asymmetry degree $\tilde{r}$ approaches zero.

To quantify this effect, we adopt the method in Ref.~\citep{fu2015effective} to compute $\tilde{r}$ for the AH-$\beta$ model. As shown in the inset of Fig.~\ref{fig5}, $\tilde{r}$ decreases approximately exponentially with increasing $\beta$. Based on this observation, we introduce a correction $\tilde{r} = r e^{-\gamma \beta}$, where $\gamma$ is a fitting parameter, and modify Eq.~(\ref{eqTeqAHB}) accordingly
 \begin{equation}\label{eqTeqAHBM}
 T_{\rm eq}^{\text{AH-}\beta} = \frac{67.17 (r e^{-\gamma \beta})^{-2.65} \beta^{-2}}{22.39 (r e^{-\gamma \beta})^{-2.65} + 3\beta^{-2}}.
 \end{equation}

The solid lines in Fig.~\ref{fig5} represent fits using Eq.~(\ref{eqTeqAHBM}). These curves closely follow the numerical data, validating the correction to a certain extent. Notably, the fitted values of $\gamma$ (e.g., $\gamma = 72.21$ for $r = 0.05$; $\gamma = 18.58$ for $r = 0.1$) differ from those derived directly from $\tilde{r}$ (3.35), and show dependence on $r$, while the original $\tilde{r}$ does not (see the inset in Fig.~\ref{fig5}). This discrepancy may arise from two approximations: the exponential form assumed for $\tilde{r}$, and the application of MR under the implicit assumption that $\tilde{r}$ and $\beta$ are independent. The precise origin of the difference remains an open question that requires further investigation.

Nevertheless, the improved agreement between theory and simulation suggests that accounting for the reduction of effective asymmetry due to nonlinearity is crucial. For fixed $r$, increasing $\beta$ suppresses the asymmetry of the IIP, leading to the observed non-monotonic behavior of $T_{\rm eq}$. It is also worth noting that, for fixed $\beta$, variations in $r$ could influence the effective strength of nonlinearity through changes in the distribution of relative displacements and energy partition. However, Fig.~\ref{fig4} shows that this effect is minimal, as the numerical data remains consistent with the uncorrected MR prediction.

\section{Conclusions}\label{sec4}

In summary, we have investigated the thermalization behavior of a 1D lattice system, focusing on the role of interaction asymmetry and nonlinearity. We first introduced the AH model, where particles interact through a purely asymmetric, non-smooth potential without explicit nonlinearity, to isolate the effects of IIP asymmetry. Numerical results show that, unlike the perturbed Toda chain, the AH model exhibits no metastable state (see Fig.~\ref{fig1}), as the non-smoothness at $x=0$ breaks Toda integrability. Thus, the AH model is best regarded as a perturbation of the harmonic chain, with the degree of asymmetry serving as the perturbation strength.

In the thermodynamic limit, we find that the thermalization time $T_{\rm eq}$ follows a power-law dependence on the perturbation strength, though with an exponent larger than the universal value reported in earlier studies (see Fig.~\ref{fig3}). This result is qualitatively consistent with findings for the FPUT-$\alpha$ (cubic) and quintic nonlinear chains~\citep{Fu_2019PRER, Pistone2019ME}. Our results—as well as those in Ref.~\citep{Fu_2019PRER}—indicate that finite-size effects are negligible in models with asymmetric IIPs, with rapid convergence observed as system size increases. Therefore, the enhanced slope is more likely due to higher-order contributions arising from asymmetry. Additionally, the AH model avoids the numerical blowup encountered in simulations of odd-order nonlinear systems~\citep{Pistone2019ME, Carati2018JSP}, making it a suitable framework for studying higher-order effects. Nevertheless, the mechanisms by which asymmetry amplifies higher-order contributions and leads to a steeper scaling exponent remain open questions.

We further examined the AH-$\beta$ model, in which asymmetry and quartic nonlinearity are simultaneously present. The thermalization time in this case is well described by MR, and in the strong nonlinearity limit, the inverse-square law for $T_{\rm eq}$ is recovered. However, a clear deviation between numerical data and the uncorrected MR prediction is observed (see Fig.~\ref{fig5}), particularly in the intermediate regime. This deviation arises from the interdependence of asymmetry and nonlinearity, which violates the independence assumption underpinning MR. Such interwoven effects often give rise to complex behaviors, highlighting the importance of isolating and characterizing individual mechanisms. The AH and AH-$\beta$ models together offer a clear and tractable framework for this purpose.

\section*{Acknowledgment}

This research was funded by the National Natural Science Foundation of China (Grants No. 12465010, No. 12247106, No. 12005156, No. 11975190, and No. 12247101). W. Fu was also supported by the Youth Talent (Team) Project of Gansu Province (No. 2024QNTD54), the Gansu Province Long-yuan Youth Talent Project, the Fei-tian Scholars Project of Gansu Province, the Leading Talent Project of Tianshui City, the Innovation Fund from the Department of Education of Gansu Province (Grant No.~2023A-106), and the Open Project Program of the Key Laboratory of Atomic and Molecular Physics $\&$ Functional Material of Gansu Province (Grant No. 6016-202404).

\bibliography{refAH}

\end{document}